# Galaxy Rest-frame UV Colors at $z \sim 2$–4 with HST UVCANDELS

Alexa M. Morales[1,10], Steven L. Finkelstein[1], Micaela B. Bagley[1], Anahita Alavi[2], Norman A. Grogin[3], Nimish P. Hathi[3], Anton M. Koekemoer[3], Kalina V. Nedkova[3,4], Laura Prichard[3], Marc Rafelski[3,4], Ben Sunnquist[3], Sina Taamoli[5], Harry I. Teplitz[2], Xin Wang[6,7,8], Rogier A. Windhorst[9], and L. Y. Aaron Yung[3]

[1] Department of Astronomy, The University of Texas at Austin, 2515 Speedway, Austin, TX 78712, USA; alexa.morales@utexas.edu
[2] IPAC, Mail Code 314-6, California Institute of Technology, 1200 E. California Blvd., Pasadena, CA 91125, USA
[3] Space Telescope Science Institute, 3700 San Martin Drive, Baltimore, MD 21218, USA
[4] Department of Physics and Astronomy, Johns Hopkins University, 3400 North Charles Street, Baltimore, MD 21218, USA
[5] Department of Physics and Astronomy, University of California, Riverside, 900 University Avenue, Riverside, CA 92521, USA
[6] School of Astronomy and Space Science, University of Chinese Academy of Sciences (UCAS), Beijing 100049, People's Republic of China
[7] Institute for Frontiers in Astronomy and Astrophysics, Beijing Normal University, Beijing 102206, People's Republic of China
[8] National Astronomical Observatories, Chinese Academy of Sciences, Beijing 100101, People's Republic of China
[9] School of Earth and Space Exploration, Arizona State University, Tempe, AZ 85287-6004, USA

## Abstract

We present an analysis of rest-frame ultraviolet (UV) colors of 17,243 galaxies at $z \sim 2$–4 in the Hubble Space Telescope UVCANDELS fields: GOODS-N, GOODS-S, COSMOS, and EGS. Here, we study the rest-frame UV spectral slope, $\beta$, measured via model spectra obtained via spectral energy distribution (SED) fitting, $\beta_{\rm SED}$, and explore its correlation with various galaxy parameters—photometric redshift, UV magnitude, stellar mass, dust attenuation, star formation rate (SFR), and specific SFR—obtained via SED fitting with DENSE BASIS. We also obtain measurements for $\beta$ via photometric power-law fitting and compare them to our SED-fit-based results, finding good agreement on average. While we find little evolution in $\beta$ with redshift from $z = 2$–4 for the full population, there are clear correlations between $\beta$ (and related parameters) when binned by stellar mass. For this sample, lower-stellar-mass galaxies ($\log[M_*] = 7.5$–$8.5\ M_\odot$) are typically bluer ($\beta_{\rm SED} = -2.0^{+0.2}_{-0.2}$ / $\beta_{\rm PL} = -2.1^{+0.4}_{-0.4}$), fainter ($M_{\rm UV} = -17.8^{+0.7}_{-0.6}$), and less dusty ($A_{\rm v} = 0.4^{+0.1}_{-0.1}$ mag), and they exhibit lower rates of star formation ($\log[{\rm SFR}]=0.1^{+0.2}_{-0.2} M_\odot\ {\rm yr}^{-1}$) and higher specific star formation rates ($\log[{\rm sSFR}] = -8.2^{+0.2}_{-0.2}\ {\rm yr}^{-1}$) than their high-mass counterparts. Higher-mass galaxies ($\log[M_*] = 10.0$–$12.0\, M_\odot$) are on average redder ($\beta_{\rm SED} = -0.9^{+0.8}_{-0.5}/\beta_{\rm PL} = -1.0^{+0.8}_{-0.5}$), brighter ($M_{\rm UV} = -19.6^{+1.0}_{-1.2}$), and dustier ($A_{\rm v} = 0.9^{+0.5}_{-0.4}$ mag), and they have higher SFRs ($\log[{\rm SFR}] = 1.2^{+0.6}_{-1.1} M_\odot\ {\rm yr}^{-1}$) and lower sSFRs ($\log[{\rm sSFR}] = -9.1^{+0.5}_{-1.1}\ {\rm yr}^{-1}$). This study's substantial sample size provides a benchmark for demonstrating that the rest-frame UV spectral slope correlates with stellar-mass-dependent galaxy characteristics at $z \sim 2$–4, a relationship less discernible with the smaller data sets typically available at higher redshifts.

## 1. Introduction

Rest-frame ultraviolet (UV) observations are crucial for identifying and analyzing young, massive stars within galaxies. These observations can constrain dust content, star formation rates (SFRs), metallicity, and the ages of stellar populations, all of which are crucial for piecing together the narrative of galaxy evolution (S. L. Finkelstein et al. 2012; A. Rogers et al. 2013; R. J. Bouwens et al. 2014; M. Castellano et al. 2014; S. de Barros et al. 2014; D. Schaerer et al. 2015; N. A. Reddy et al. 2018; A. Calabrò et al. 2021). These properties significantly influence a galaxy's rest-frame UV color, a critical marker of its underlying physical processes. The UV spectral slope, $\beta$, is especially informative, reflecting the contour of a galaxy's UV continuum and serving as a proxy for its intrinsic UV luminosity, among other characteristics (D. Calzetti et al. 1994; G. R. Meurer et al. 1999). Studying the UV continuum and its linkage with $\beta$ allows for a more nuanced understanding of the galaxy's youth, its star-forming activity, and the presence of dust, all of which leave an imprint on its UV light profile.

Recent studies leveraging Hubble Space Telescope (HST) and James Webb Space Telescope (JWST) observations have expanded our understanding of UV spectral slopes in star-forming galaxies at $z \geqslant 4$ (S. L. Finkelstein et al. 2012; R. J. Bouwens et al. 2014; R. Bhatawdekar & C. J. Conselice 2021; M. W. Topping et al. 2022, 2024; D. Austin et al. 2023; F. Cullen et al. 2023; A. M. Morales et al. 2024; X. Wang et al. 2024a). These empirical results from photometric analyses have elucidated a clear trend: a progression toward bluer UV spectral slopes with increasing redshift, indicating galaxies characterized by intense star formation, younger stellar populations, and minimal dust attenuation. In addition to redshift, previous works have underscored correlations of $\beta$ with stellar mass (S. L. Finkelstein et al. 2010, 2012; N. P. Hathi et al. 2013), UV magnitude (R. J. Bouwens et al. 2010, 2012; N. P. Hathi et al. 2016), and dust attenuation (D. Calzetti et al. 1994; G. R. Meurer et al. 1999). These relationships are physically motivated, as UV magnitude and stellar mass are interconnected through star formation processes, with more massive galaxies typically exhibiting higher UV luminosities. Dust attenuation further complicates this relationship by obscuring and scattering UV light, thereby altering the observed characteristics of galaxies.

[10] NSF Graduate Research Fellow.

**Table 1**
List of Filters Utilized for This Analysis for Each UVCANDELS Field: GOODS-N, GOODS-S, EGS, and COSMOS

| Field | Telescope: Instrument | Filter |
| --- | --- | --- |
| GOODS-N | HST: WFC3/ACS | F275W, F435W, F606W, F775W, F850LP, F105W, F125W, F140W, F160W |
| | KPNO: Mosaic | *U* band |
| | LBT: LBC | *U* band |
| | Spitzer: IRAC | Channel 1, Channel 2 |
| GOODS-S | HST: WFC3/ACS | F275W, F435W, F606W, F775W, F814W, F850LP, F098M, F105W, F125W, F160W |
| | Paranal: VIMOS | *U* band |
| | Spitzer: IRAC | Channel 1, Channel 2 |
| EGS | HST: WFC3/ACS | F275W, F435W, F606W, F814W, F125W, F140W, F160W |
| | Spitzer: IRAC | Channel 1, Channel 2 |
| COSMOS | HST: WFC3/ACS | F275W, F435W, F606W, F814W, F125W, F160W |
| | CFHT-LS: Megaprime | $u^*$ band, $g^*$ band, $r^*$ band, $i^*$ band, $z^*$ band |
| | Subaru: Suprime | IAL-527, *g*' band, IB-624, IB-679, *V* band, *r*' band, NB-711, IB-738, IB-767, *i*' band, *z*' band, NB-816 |
| | NOAO: NEWFIRM | *J*1 band, *J*2 band, *J*3 band, *H*1 band, *H*2 band |
| | Paranal: VISTA | *Y* band, *J* band, *H* band, *Ks* band |
| | Spitzer: IRAC | Channel 1, Channel 2 |

Building upon this high-redshift foundation, the present work shifts the focus to the less explored intermediate redshift range of $z \sim 2$–4, where previous analyses have been limited by shallower depths or more restricted sky coverage (N. P. Hathi et al. 2013, 2016). The sensitive space-based UV data now available from HST's Wide Field Camera 3 (WFC3; M. Stiavelli & R. O'Connell 2001) and the Advanced Camera for Surveys (ACS; J. E. Ryon & D. V. Stark 2023), in particular through the UVCANDELS survey (PI: Teplitz), opens up new avenues for research. UVCANDELS provides high-resolution UV imaging across four well-studied fields (GOODS-N, GOODS-S, COSMOS, and EGS), significantly enhancing the depth and sky coverage compared to earlier works. When combined with rest-frame optical data from Spitzer/IRAC and the comprehensive HST CANDELS survey (Co-PIs: Faber & Ferguson; N. A. Grogin et al. 2011; A. M. Koekemoer et al. 2011), this data set enables a more robust analysis of UV slopes and galaxy properties such as stellar mass, dust attenuation, and star formation rates. The improved depth allows us to probe fainter galaxies, in particular those with lower stellar masses, while the wider sky coverage mitigates the effects of cosmic variance. This advancement enables us to explore the relationship between UV slopes and galaxy parameters with greater precision and over a wider range of stellar masses and redshifts than previously possible, addressing key limitations of prior studies.

Beyond the isolated examination of galaxies at $z = 2$–4 as a whole, given a large enough sample such as this one, a more interesting perspective emerges when galaxies are dissected based on their stellar mass, acknowledging that galaxies with varying stellar masses may follow distinct evolutionary trajectories. This approach allows us to investigate the UV spectral slope as a function of stellar mass, offering a potent tool for untangling the complex physical processes that underlie the observed diversity of galaxies during this epoch. Understanding how the rest-frame UV spectral slope varies with stellar mass can shed light on the interplay between star formation, dust content, and other galaxy properties. It allows us to explore whether lower-mass galaxies exhibit different trends in UV color compared to their higher-mass counterparts. Such insights can deepen our understanding of how galaxies of different sizes and mass assemble their stellar populations and evolve over cosmic time. The examination of galaxy colors at $z = 2$–4 also sets a foundation for understanding galaxy evolution at even higher redshifts. As observations are extending our observations into the Epoch of Reionization and beyond, using facilities like JWST, the lessons learned from HST studies serve as benchmarks. This continuity is vital for understanding the complex processes that govern galaxy formation and evolution across cosmic time.

This paper is structured as follows. In Section 2, we describe the UVCANDELS survey and the data reduction process. We also discuss our process for obtaining our sample and information returned from photometric power-law and spectral energy distribution (SED) fitting. In Section 3, we describe our findings from fitting observations and simulations to models and ties to galaxy parameters. In Section 4, we discuss our results, and we present our conclusions in Section 5.

We use the Planck Collaboration et al. (2020) cosmology of $H_0 = 67.4\ \mathrm{km\ s^{-1}\ Mpc^{-1}}$, $\Omega_m = 0.315$ and $\Omega_\Lambda = 0.685$, and all magnitudes are given in the AB system (J. B. Oke & J. E. Gunn 1983).

## 2. Methods

In Section 2.1, we briefly describe the UVCANDELS survey and its data reduction process. In Section 2.2, we describe the methodology used to select galaxies at $z \sim 2$–4. Section 2.3 describes our SED-fitting process with DENSE BASIS to get $\beta_{\rm SED}$, and Section 2.4 describes the photometric power-law fitting applied to these sources to obtain $\beta_{\rm PL}$.

### *2.1. Data*

We select the sample of galaxies from four of the Cosmic Assembly Near-Infrared Deep Extragalactic Survey (CANDELS; Co-PIs: Faber & Ferguson; N. A. Grogin et al. 2011; A. M. Koekemoer et al. 2011) fields that were also observed by the UVCANDELS survey (PI: Teplitz; X. Wang et al. 2024b). UVCANDELS covers 430 arcmin$^2$ with WFC3-UVIS F275W, and approximately the same area with ACS F435W. The four fields and the corresponding photometric filters that were used for this analysis are listed in Table 1. The UV images reach about 27th magnitude (AB, 5$\sigma$) for compact sources (measured in a 0″.2 radius). The F435W data were taken in parallel and reached about 28th magnitude (AB, 5$\sigma$). UVCANDELS provides the first wide-area F435W observations in COSMOS and EGS. The GOODS fields have previous imaging in the same filter, so the new imaging was placed in the CANDELS-deep regions, where other archival UV data are also available.

The reductions of UVCANDELS images are described in detail in X. Wang et al. (2024b), so we only briefly summarize the process here. Calibration of the images has been improved from the standard products available at the time, including custom darks with improved hot pixel rejection, custom cosmic-ray and readout cosmic-ray rejection, equalization of amplifier background levels, and removal of gradients from scattered light in the ACS images, using methods developed by M. Rafelski et al. (2015), L. J. Prichard et al. (2022), and M. Revalski et al. (2023). Individual exposures have been astrometrically aligned, and image stacks have been created using the pipeline developed by A. Alavi et al. (2014). The final mosaics drizzled to match the pixel scale of the CANDELS mosaics (30 and 60 mas) are available at the Mikulski Archive for Space Telescopes (MAST).[11]

The creation of photometric catalogs for UVCANDELS is presented in L. Sun et al. (2024). Briefly, they use a method of UV-optimized aperture photometry developed by M. Rafelski et al. (2015). In this method, isophotes are defined in an optical CANDELS F606W band and used to measure the signal in the UV. These isophotes are better matched to the sizes of sources detected in the UV images, and so they reduce noise that would be introduced by using $H$-band apertures (e.g., G. Barro et al. 2019). PSF and aperture corrections are applied at the catalog level by comparison with the CANDELS F606W catalogs. The UVCANDELS catalogs (the recent paper of X. Wang et al. 2024b details the methodologies and data discussed) will be available in MAST in mid-2024.

### 2.2. Sample Selection

To obtain our initial sample, for each field, we run the entire catalog through EAZY (G. B. Brammer et al. 2010), whereby we obtain the redshift probability distribution, $P(z)$, and best-fit redshift, $z_a$. We use the default set of 12 "tweak FSPS" templates and included six additional templates from R. L. Larson et al. (2023), which provide greater flexibility in fitting a broader range of SED shapes, including bluer colors. These additional templates allow us to better capture the diversity of galaxies in our sample, in particular those that may deviate from the typical SED shapes of high-redshift galaxies, without forcing bluer colors.

We assume a flat redshift prior with respect to luminosity and allow the redshift to range from $z = 0$–15. No redshift or magnitude priors are imposed during the fitting process, ensuring the results are not biased by prior assumptions. Redshift and magnitude cuts are applied only after the fitting has been completed.

We then put sources through a detailed selection process that evaluates whether they are viable for further analysis. Sources in each of the four fields must satisfy the following criteria:

1. The redshift probability distributions, $P(z)$, must have the majority of the integrated distribution, i.e., >50%, fall within a specific redshift range of our choice. For $z = 2$, this is $1.5 \leqslant z_a \leqslant 2.5$, for $z = 3$, it is $2.5 < z_a \leqslant 3.5$, and for $z = 4$, it is $3.5 < z_a \leqslant 4.5$.
2. Signal-to-noise ratios (SNRs) in both the $J_{125}$ and $H_{160}$ imaging bands $\geqslant 3.5$, or $\geqslant 5$ in either band, to ensure that the source is significantly detected and not spurious.
3. $\chi^2_{\rm EAZY} \leqslant 50$ (to ensure a reasonable fit to the data).
4. A magnitude cutoff in the $H_{160}$ filter band $\geqslant 20$, to remove possible stars, following S. L. Finkelstein et al. (2012).

To improve the alignment of the EAZY results with the observations, we derive a zero-point offset correction on a per-filter basis. We do this using sources in each catalog that have published spectroscopic redshifts (spec-$z$; catalog obtained from N. Hathi 2025, private communication). We rerun EAZY on these sources with the redshift fixed to the spec-$z$ and take the ratio of the model to observed flux densities in each filter bandpass as a correction factor. We apply the median ratio to each filter and iterate this process again (three times in total) until the median ratio of the model to observed fluxes is ∼1. The final zero-point correction is the product of all three iterations, which we then apply to each filter for the entirety of the sample in each field. However, it is important to acknowledge that this sample is very heterogeneous, which might introduce biases that are not easily quantified; thus, these per-filter offsets might not uniformly correct for different categories of sources, in particular those not well represented by the published spec-$z$ sample. With the zero-point corrections applied to each field's photometric table, we rerun EAZY, reapply our selection process to the full catalog, and obtain our final sample.

To ensure our criteria are retrieving galaxies accurately, we visually inspect 1000 sources in each field, wherein we look at their image stamps in each HST filter, alongside their corresponding EAZY $P(z)$ distribution and SED fit to the observed photometry. We compare the best-fit template photometry and SED at the minimum chi-square photometric redshift to the corrected input photometry. On average, the agreement between the template and the input photometry is reliable, with no significant systematic offsets observed. We evaluate whether or not each source is a "true" galaxy, in which we see a clear dropout in the corresponding filter where the redshift was estimated to be, the image stamps show a clear source in all of the detection bands, and the SED is a good fit to the observed photometry. This methodological approach not only ensures the accuracy of our galaxy selection but also enhances the efficiency of the process, negating the need for an impractical line-by-line inspection of all ∼17,000 candidates in the data set. Notably, only about 1% of the visually inspected plots have been deemed questionable, affirming the robustness of our selection criteria.

Objects that satisfy the selection criteria with EAZY are then fit with DENSE BASIS. DENSE BASIS is a Bayesian SED-fitting code that utilizes nonparametric star formation histories represented by a Gaussian mixture model (K. G. Iyer et al. 2019) and stellar templates provided by FSPS (C. Conroy et al. 2009). DENSE BASIS is specifically designed to handle large data sets efficiently while maintaining the flexibility of nonparametric SFH reconstruction. The nonparametric SFHs allow for a flexible representation of galaxy evolution,

[11] https://archive.stsci.edu/hlsp/uvcandels

**Table 2**
Total Number of Sources, $N$, in Each UVCANDELS Field Utilized for This Work (GOODS-N, GOODS-S, COSMOS, and EGS) for $z = 2$–4

| Field | $N_{z=2}$ | $N_{z=3}$ | $N_{z=4}$ | Area (arcmin$^2$) | 5$\sigma$ Depth in F160W (mag) |
|---|---|---|---|---|---|
| GOODS-N | 3136 | 1107 | 302 | 171 | 27.80 |
| GOODS-S | 1296 | 1013 | 321 | 170 | 27.36 |
| COSMOS | 1555 | 576 | 179 | 216 | 27.56 |
| EGS | 4882 | 2209 | 680 | 206 | 26.62 |
| Total | 10869 | 4905 | 1482 | … | … |

**Note.** Here, the $z = 2$ bin spans $1.5 < z \leqslant 2.5$, $z = 3$ spans $2.5 < z \leqslant 3.5$, and $z = 4$ spans $3.5 < z \leqslant 4.5$.

capturing complex star formation scenarios that fixed parametric forms (e.g., exponentially declining or delayed SFHs) cannot. This flexibility is particularly important for accurately modeling galaxies at high redshifts, where star formation histories are often more stochastic. We generate a mock stellar population with FSPS (using `fsps.StellarPopulation()`), where we implement a G. Chabrier (2003) initial mass function (IMF) and a D. Calzetti et al. (2000) dust curve. Here, photometric data are fed into an "atlas," where a list of corresponding filter curves and a suite of priors are defined (Table 3). These priors span a wide range of values for metallicity, dust, and specific star formation rate, and are flat in nature to provide an extensive range of SED shapes and corresponding galaxy parameters to be fit to the photometry. We also incorporate a prior on stellar mass, the redshift range tested, and a redshift prior, which takes the redshift probability distribution we obtained from EAZY. When utilizing the redshift probability distribution function, following the work of K. Chworowsky et al. (2024), we modify and reshape our EAZY $P(z)$ to fit a top-hat function, because the prior DENSE BASIS assumed redshifts can only be in a top-hat form. This approach ensures consistency between the redshifts derived from EAZY and DENSE BASIS. In practice, we find that the best-fit redshifts from DENSE BASIS and EAZY are in excellent agreement, with a mean deviation of $-0.022$ and a scatter of 0.121. To further validate our redshifts, we compare the DENSE BASIS photometric redshifts to the spectroscopic redshifts available for a subset of sources. The median values of $\Delta z/(1 + z)$ are very close to zero ($\sim$0.01–0.03) in all fields, indicating good agreement with the spectroscopic redshifts.

The best-fit SED and corresponding galaxy parameter posteriors are returned as a result. We assess both the likelihood of the model SEDs and the associated galaxy parameters. DENSE BASIS, by default, retains the top 100 results with the highest likelihood for further analysis. From there, we end up with a final sample of 17,243 galaxies across all four UVCANDELS fields. See Table 2 for a breakdown of the number of sources in each field.

### 2.3. Measuring β via SED Fitting with DENSE BASIS

We aim to measure $\beta$ from these model spectra directly, following S. L. Finkelstein et al. (2012). DENSE BASIS creates an atlas of model spectra based on the defined priors (see Section 2.2 and Table 3 for information on DENSE BASIS). It retains the top 100 results with the highest likelihood for further analysis.

**Table 3**
DENSE BASIS Atlas Priors Defined for This Work

| Parameter | Range | Description |
|---|---|---|
| **Metallicity Component** | | |
| Metallicity ($Z$) range | (−2.0, 0.25) | Metallicity in units of solar metallicity, $Z_\odot$ |
| Prior | Flat | |
| **Dust Component** | | |
| Shape | Calzetti | Shape of the attenuation curve |
| Dust attenuation ($A_v$) range | (0.0, 4.0) | Dust attenuation in units of magnitude |
| Prior | Flat | |
| **sSFR Component** | | |
| log(sSFR) range | (−14.0, −7.0) | sSFR in units of yr$^{-1}$ |
| Prior | Flat | |
| **Additional Fit Instructions** | | |
| Redshift ($z$) range | (0.0, 6.0) | Redshift range tested |
| Stellar mass ($\log(M_*)$) range | (6.0, 12.5) | Stellar mass in units of solar mass, $M_\odot$ |

From the 100 measurements for each source, we compute the median values and uncertainties for various galaxy properties, including redshift, stellar mass, UV magnitude (derived from the bandpass average flux at rest-frame 1500 Å), dust attenuation, star formation rate, and specific star formation rate. These values are determined by calculating the median and 1$\sigma$ spread.

We also utilize the median (and the difference between the median and the 68% confidence bounds) for redshift, stellar mass, UV magnitude, dust attenuation, star formation rate, and specific star formation rate as our final values (and error bars) in this work. For each source, DENSE BASIS fits model SEDs to the photometric data points and their errors, returns the best-fit SED model, and provides 100 draws from the resulting posteriors for various galaxy properties as a function of these SEDs.

After assessing the likelihood and associated parameters of these models, we utilize all of the flux density data points from the spectra for each posterior SED model with DENSE BASIS, as defined by D. Calzetti et al. (1994), to measure $\beta$ directly, following S. L. Finkelstein et al. (2012). With these data points, we fit a power law to all of the points where $\log(f_\lambda) = \beta \log(\lambda) + \log(y_{\rm int})$. We then take the median and difference from the 1$\sigma$ bounds for the 100 measurements of $\beta_{\rm SED}$ as our final value and error bars.

### 2.4. Measuring β via Photometric Power-law Fitting with EMCEE

Here, we describe our process of measuring the UV spectral slope via photometric power-law fitting to the observed photometry. Measuring the UV spectral slope in this manner is not reliant on stellar population models, though it is dependent on the number of photometric data points available. When measuring the UV spectral slope with photometric power-law fitting to the observed photometry, $\beta_{\rm PL}$, we redshift the rest-frame $\lambda = 1500$–3000 Å regime to the corresponding median redshift estimated from EAZY, $z_a$, for each galaxy in the sample, and we keep the filters whose filter curves fall fully within the redshifted wavelength range. Once the photometric data points within the wavelength range are determined, we fit the data points to a line (as defined at the end of Section 2.3) and run this process through EMCEE (D. Foreman-Mackey et al. 2013). This procedure maximizes the likelihood that the model described by three free parameters matches the observed photometry for a given source: $\beta$, $y_{\rm int}$, and a fractional error factor, $\log(f)$, where it is assumed that the likelihood function and its uncertainties are simply a Gaussian, where the variance is underestimated by some fractional amount. Results are derived from the median and 68th percentile of the posterior distribution on these three parameters from a chain consisting of 5000 steps. To maintain uniformity across how $\beta_{\rm PL}$ is defined, we omit 531 sources from this specific figure, due to the lack of data points allowable to measure the UV slope (i.e., these sources only have one filter that lies within the wavelength range we set).

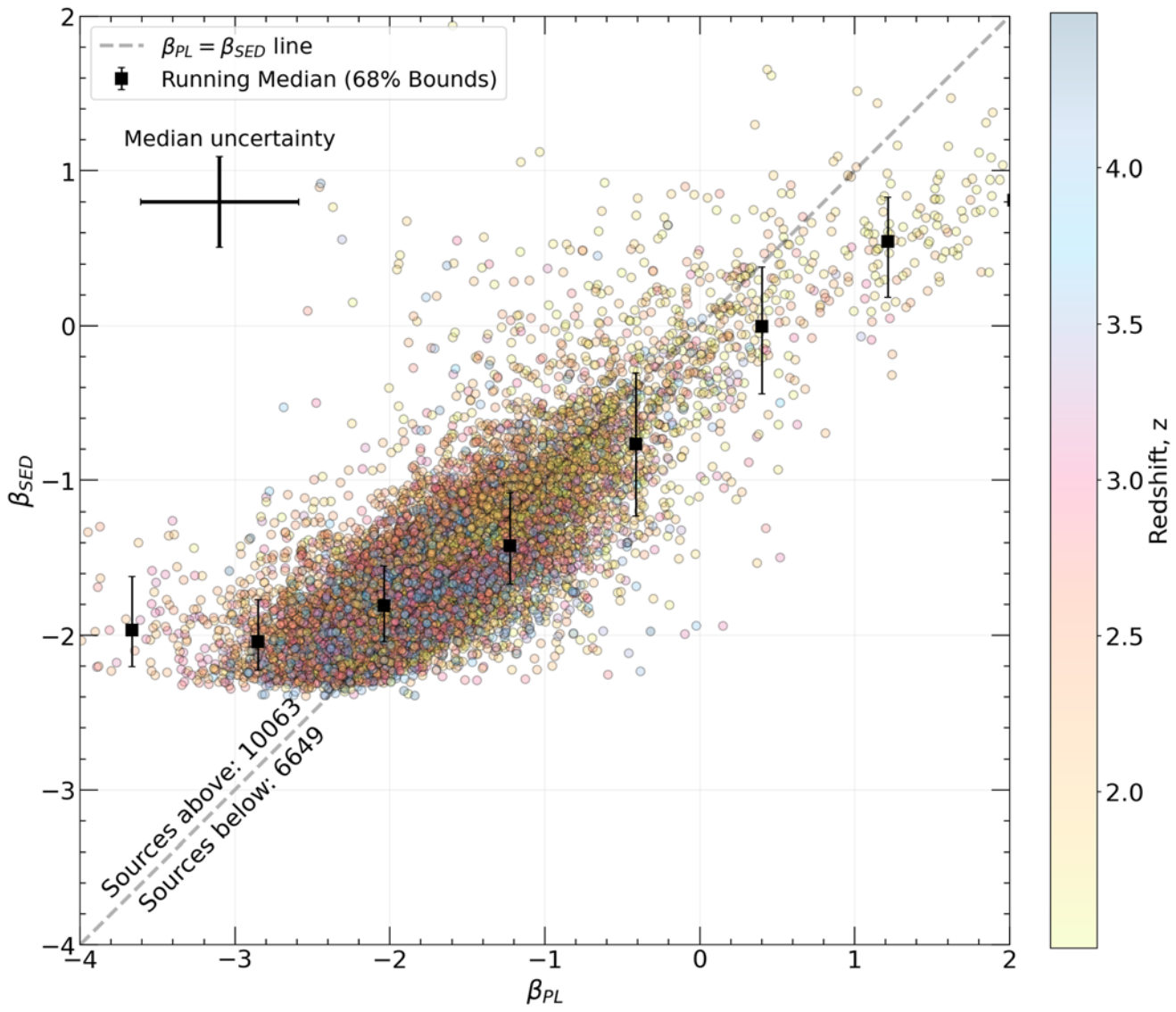

**Figure 1.** UV $\beta$ slope derived from photometric power-law fitting vs. that from SED fitting with DENSE BASIS. The sample is color-coded by redshift to show the distribution of galaxies at $z = 2$–4 and their corresponding UV colors. The gray dashed line is the one-to-one line wherein we also list the number of sources above and below this line. The median uncertainty for each $\beta$ is derived from the upper and lower error bars ($1\sigma$ difference from the median of each galaxy's beta distribution, shown as black squares), where $\sum\sigma_{\rm up} + \sum\sigma_{\rm low}/2$. This relation shows that photometric power-law fitting, on average, yields bluer UV slopes and larger errors than SED fitting.

### 2.5. Comparison of $\beta$ from SED-fitting and Power-law Methods

UV spectral slopes for our sample from both measurement methods are shown in Figure 1. We show that the SED-fitting method yields smaller error bars on average—this is simply due to the utilization of more data points across the specified wavelength range mentioned in Section 2.3, given that the SED is a good fit to the data. This figure clearly demonstrates the density of data points around the one-to-one line ($\beta_{\rm PL} = \beta_{\rm SED}$), illustrating a good overall agreement between the power-law and SED-fitted slopes. However, a slight systematic deviation indicates that the SED-measured UV slopes are, on average, redder than those derived from photometric power-law fitting. Here, the average uncertainties for SED fitting are also approximately twice as small as those for photometric power-law fitting. While no $\beta_{\rm SED}$ values reach $\beta_{\rm SED} < -3$, there are a small number (158 sources) whose $\beta_{\rm PL} < -3$. Similarly, when looking at the red end of the distribution of $\beta$, while few reach $\beta_{\rm SED} > 1$ (21 sources), there are 105 sources whose $\beta_{\rm PL} > 1$. Although the default stellar grids used in DENSE BASIS provide sufficient grounds to do this analysis, there will always be a bias with regard to what is possibly excluded. For instance, the stellar grids utilized by default with DENSE BASIS can lead a $\beta$ "floor," where the UV slope $\beta$ does not realistically extend below approximately $-2.5$, potentially skewing the SED fits for galaxies with extreme properties (see Section 4.1 for more information). However, we note that the majority of our sample and their relative uncertainties sit well above this floor. That said, since no combination of SED-fitting codes or stellar templates can consistently reproduce the extreme blue end of $\beta_{\rm PL}$ we see in Figure 1. The only exception might be the E. Zackrisson et al. (2011) models, which include Population III stars with $\beta \sim -3$, though such sources are unlikely to be present in this data set. Instead, they are most likely artifacts of the photometric uncertainties and number of data points available to fit, rather than true rest-UV properties. In these cases, SED-fitting—or ideally, direct spectral observations—would provide a more reliable approach.

## 3. Results

### 3.1. Analysis of $\beta_{\rm SED}$ and Other Dense Basis Galaxy Parameters

We investigate any correlations between $\beta_{\rm SED}$ and other DENSE BASIS galaxy properties such as best-fit redshift, UV magnitude, stellar mass, dust attenuation, star formation rate, and specific star formation rate. We identify monotonic trends using Spearman R correlation coefficients, $\rho$.[12] We perform a Monte Carlo resampling to estimate the Spearman correlation coefficient and the corresponding $p$-value between $\beta_{\rm SED}$ and other galaxy parameters, accounting for their uncertainties (repeatedly adding normally distributed random noise to the data and recalculating the $\rho$ and $p$-value for the entire sample).

Previous works have mentioned correlations of $\beta$ mainly with stellar mass (S. L. Finkelstein et al. 2010, 2012; N. P. Hathi et al. 2013), UV magnitude (R. J. Bouwens et al. 2010, 2012; N. P. Hathi et al. 2016), and dust attenuation (D. Calzetti et al. 1994; G. R. Meurer et al. 1999). These correlations are physically motivated by the intrinsic properties of galaxies: UV magnitude and stellar mass are linked through the star formation processes where more massive galaxies exhibit higher UV luminosities. Dust attenuation directly affects UV magnitude by obscuring and scattering UV light, altering the observed characteristics of galaxies.

In Figure 2, we show $\beta_{\rm SED}$ plotted as a function of different galaxy properties for the full sample (without splitting by redshift). We find that dust attenuation and stellar mass exhibit the strongest (but still moderately) positive Spearman R correlations, where $A_{\rm v}$ has a $\rho = 0.41 \pm 0.02$ and a $p$-value of $0.0 \pm 0.0$, and stellar mass has a $\rho = 0.56 \pm 0.00$ and a $p$-value of $0.0 \pm 0.0$, implying that higher magnitudes of dust attenuation along the line of sight lead to redder UV slopes (and vice versa; see D. Calzetti et al. 1994; G. R. Meurer et al. 1999) and higher stellar mass also contributes to redder UV slopes. Specific star formation rate has a weak negative monotonic correlation with the UV spectral slope, with a $\rho = -0.38^{+0.00}_{-0.01}$ and a $p$-value of $0.0 \pm 0.0$, but from the figure, we can see that bluer galaxies are typically exhibiting more active star formation rates and are typically younger (R. J. Bouwens et al. 2009). Star formation rate, redshift, and UV magnitude similarly have weak to negligible trends with the UV spectral slope, with $|\rho| < 0.3 \pm 0.08$.

[12] We define correlation strengths as the following: (1) negligible $= 0.00 < |\rho| < 0.20$, (2) weak $= 0.21 < |\rho| < 0.40$, (3) moderate $= 0.41 < |\rho| < 0.60$, (4) strong $= 0.61 < |\rho| < 0.80$, and (5) very strong $= 0.81 < |\rho| < 1.00$. We also note statistical significance in the correlations, as defined by $p$-value: (1) significant $= p$-value $\leqslant 0.05$ and (2) nonsignificant $= p$-value $> 0.05$.

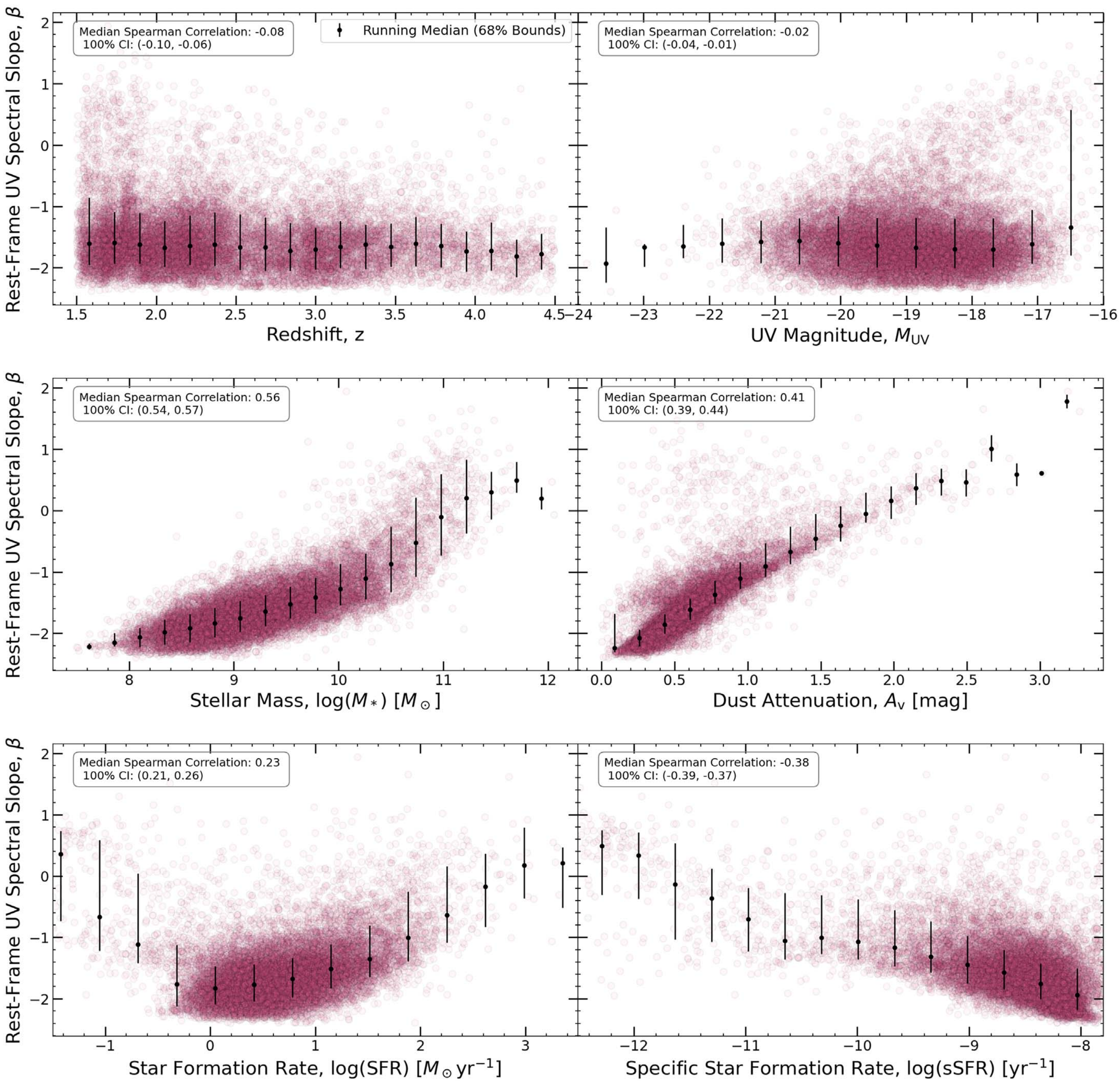

**Figure 2.** UV spectral slope, $\beta$, as a function of redshift, UV magnitude, stellar mass, dust attenuation, star formation rate, and specific star formation rate. The pink points are the individual objects from all four fields, and the black points indicate the running median, with the uncertainty being the difference from the 68th percentile. The median Spearman R correlation coefficient and $1\sigma$ confidence interval listed are with respect to the entire sample. We find little evolution in $\beta$ with redshift, UV magnitude, and star formation rate. Strikingly, there are clear trends with stellar mass and dust attenuation, wherein objects with higher stellar masses have larger magnitudes of dust attenuation and thus redder UV colors than their counterparts. Concurrently, a moderate trend emerges with specific star formation rates, with higher sSFRs linked to bluer UV colors, which are characteristic of galaxies that are more active and possess lesser stellar mass and dust attenuation.

### 3.2. Analysis of $\beta_{SED}$ as a Function of Stellar Mass

Given that strong trends were shown between the UV spectral slope and stellar mass, here we bin the results by stellar mass (we discuss this lack of observed correlation between $\beta$ and $M_{UV}$ in Section 3.1). Although the main filter set used for this analysis rests in the rest-UV regime, with the inclusion of Spitzer/IRAC (G. G. Fazio et al. 2004) filter Channels 1 and 2, we are able to obtain more robust measurements of stellar masses with SED-fitting. As such, in Figures 3, 4, and 5, we show $\beta_{SED}$ plotted as a function of each galaxy property but binned by stellar mass and redshift. Clearer trends are now observed, where for almost every galaxy property, galaxies with a smaller stellar mass exhibit bluer UV colors, are on average fainter, contain less dust, have lower star formation rates, and exhibit higher specific star formation rates than their redder, higher-mass counterparts. In Table 4, we give the median and 1$\sigma$ difference for each galaxy property binned by stellar mass and redshift (Table 5 shows median and 1$\sigma$ values for $\beta$ and galaxy properties as a function of redshift and as a whole, alongside their corresponding Spearman correlation coefficients).

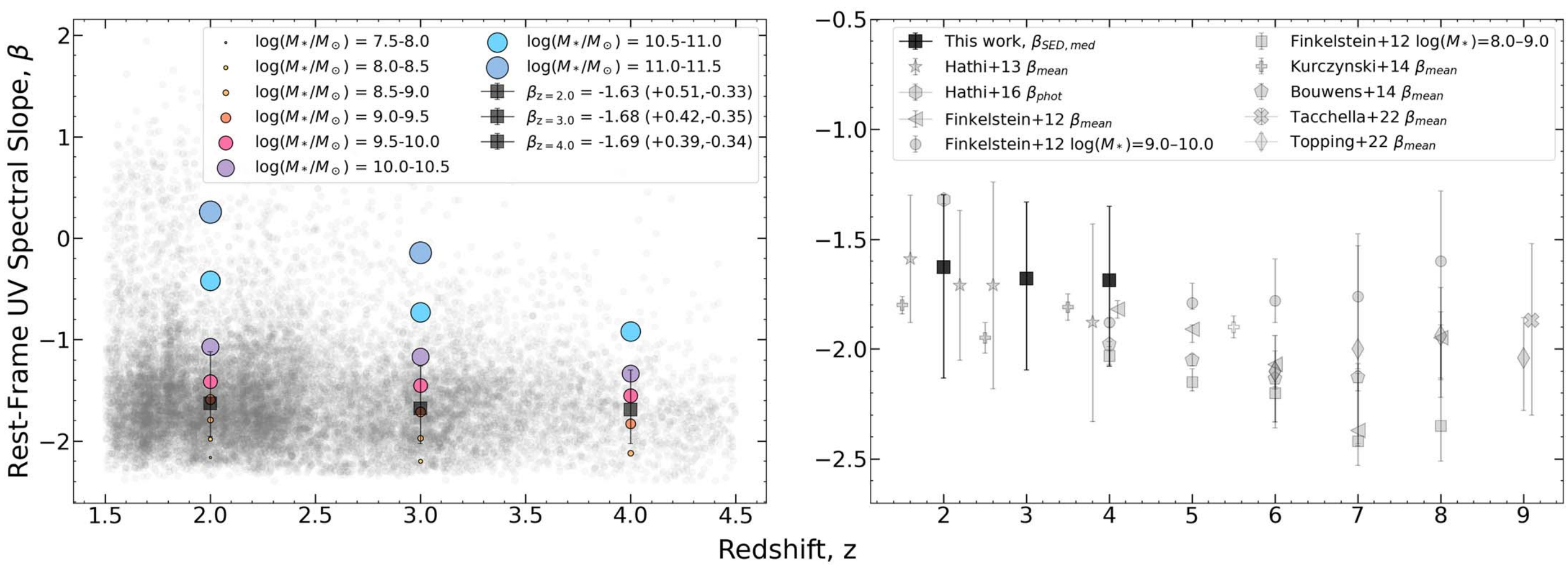

**Figure 3.** Left: we examine the UV spectral slope, $\beta$, as a function of redshift, represented by light gray circles. The median and 1$\sigma$ spread of $\beta$ for each redshift are depicted by black squares. Our analysis reveals that the UV spectral slope shows no strong evolution solely with redshift, as the average $\beta$ remains relatively constant. Additionally, when $\beta$ is binned by stellar mass (color-coded and with marker size categorized by stellar mass), it becomes evident that higher-mass sources exhibit redder UV colors compared to their lower-mass counterparts. Right: average UV spectral slope as a function of redshift, and their uncertainties compared to other observations. Our results are in general agreement with other observations at these same redshifts, and overall, the median UV slope as a function of redshift is decreasing toward bluer UV slopes. Observations are denoted as follows: This work (black squares), N. P. Hathi et al. (2013, gray stars), N. P. Hathi et al. (2016, gray hexagons), S. L. Finkelstein et al. (2012) average (gray triangles), S. L. Finkelstein et al. (2012) binned by stellar mass (log($M_*$) = 8–9, gray squares; log ($M_*$) = 9–10, gray circles), R. J. Bouwens et al. (2014, gray pentagons), S. Tacchella et al. (2022, gray x-markers), P. Kurczynski et al. (2014, gray plus signs), and M. W. Topping et al. (2022, gray diamonds).

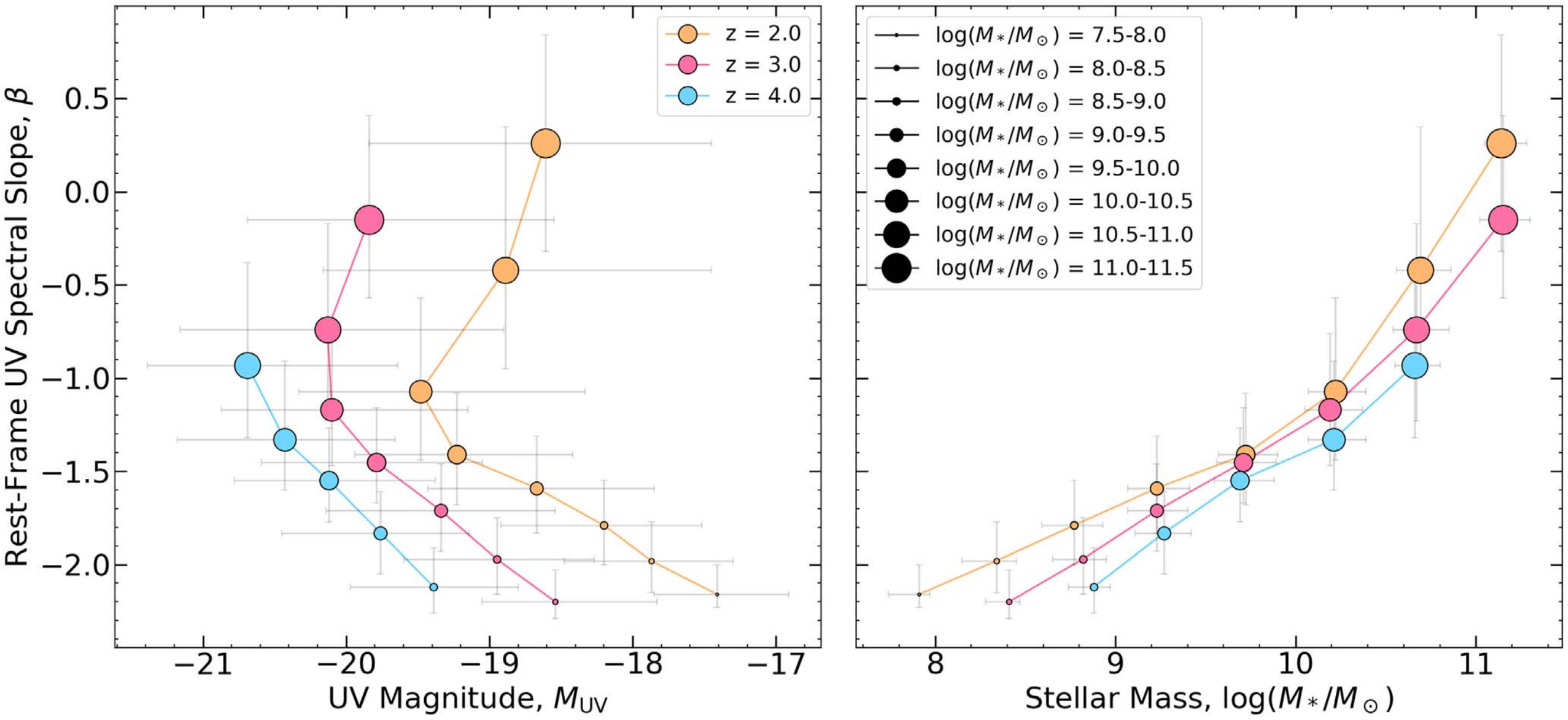

**Figure 4.** UV $\beta$ slope as a function of UV magnitude (left) and stellar mass (right), where the data are color-coded by redshift and the marker size indicates the stellar mass bin. For each redshift, we see, on average, that lower-mass sources exhibit bluer UV colors overall. These lower-mass sources exhibit brighter UV magnitudes and bluer UV colors as a function of redshift and stellar mass. The turnover in UV magnitude occurring at $z = 2$, 3 for higher-mass sources can be attributed to higher dust content lowering the overall UV magnitude and brightness (albeit with a larger spread in the 1$\sigma$ distribution when compared to lower-mass averages).

In Figure 4 ($\beta$ versus $M_{UV}$ and stellar mass) and Figure 5 ($\beta$ versus $A_v$, SFR, and sSFR), we analyze the UV slope, $\beta$, across different redshifts and stellar masses. These observations reveal several key trends:

1. *Redshift variation.* For a given stellar mass, galaxies at $z = 2$ exhibit fainter UV magnitudes on average than sources at higher redshifts in this sample. As redshift increases to $z = 4$, the sample predominantly consists of brighter galaxies with bluer UV colors compared to their $z = 2$ counterparts. This shift toward brighter, bluer galaxies at higher redshifts may reflect a selection bias or intrinsic evolutionary trends.

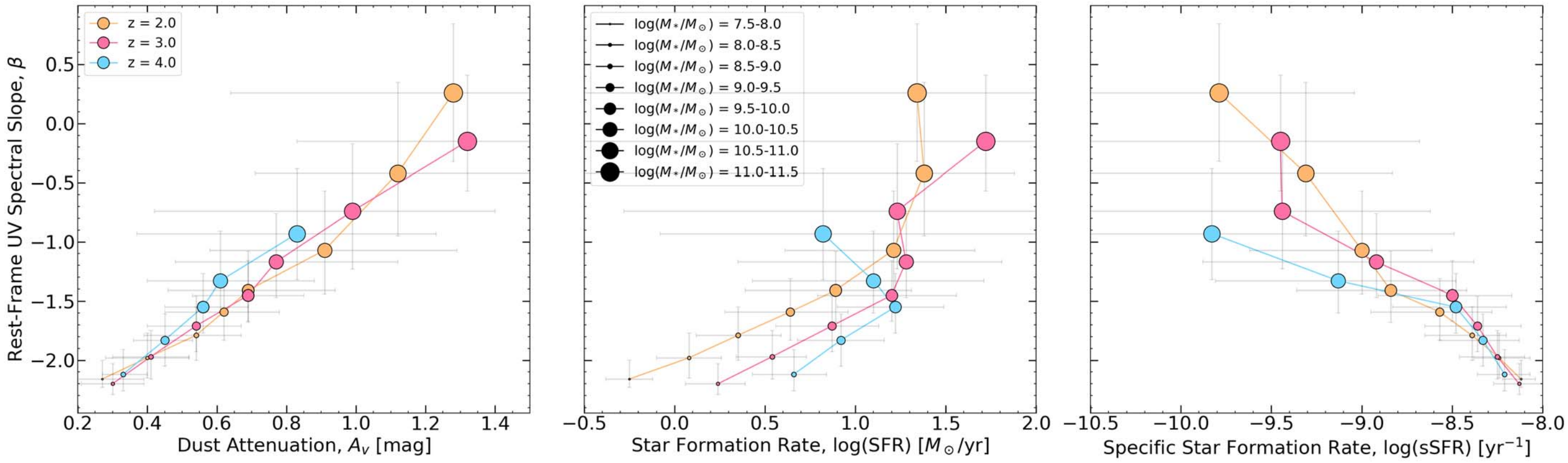

**Figure 5.** UV $\beta$ slope as a function of dust attenuation, star formation rate, and specific star formation rate (color-coded by redshift, with marker size binned by stellar mass). As discussed in Section 3.1, the galaxy parameters shown here have more obvious trends with the UV spectral slope. Namely, dust attenuation shows the strongest positive correlation with $\beta$. SFR and sSFR follow and exhibit moderate correlations with $\beta$. The corresponding Spearman R correlation coefficients are listed for each parameter in Table 5.

**Table 4**
DENSE BASIS Galaxy Properties for Stellar Mass Bins with $N > 20$

| | $\log(M_*/M_\odot)$ Bin | $N$ | $\beta_{\rm SED}$ | $M_{\rm UV}$ | $A_{\rm v}$ | log(SFR) | log(sSFR) |
|---|---|---|---|---|---|---|---|
| | 7.5–8.0 | 123 | $-2.16^{+0.16}_{-0.07}$ | $-17.41^{+0.50}_{-0.44}$ | $0.27^{+0.12}_{-0.07}$ | $-0.25^{+0.13}_{-0.13}$ | $-8.12^{+0.10}_{-0.08}$ |
| | 8.0–8.5 | 1482 | $-1.98^{+0.21}_{-0.17}$ | $-17.87^{+0.57}_{-0.61}$ | $0.40^{+0.12}_{-0.12}$ | $0.08^{+0.18}_{-0.18}$ | $-8.24^{+0.17}_{-0.14}$ |
| | 8.5–9.0 | 2925 | $-1.79^{+0.24}_{-0.21}$ | $-18.20^{+0.68}_{-0.72}$ | $0.54^{+0.13}_{-0.15}$ | $0.35^{+0.24}_{-0.23}$ | $-8.39^{+0.19}_{-0.26}$ |
| $z = 2$ | 9.0–9.5 | 2712 | $-1.59^{+0.28}_{-0.24}$ | $-18.67^{+0.82}_{-0.76}$ | $0.62^{+0.16}_{-0.17}$ | $0.64^{+0.32}_{-0.36}$ | $-8.57^{+0.29}_{-0.34}$ |
| | 9.5–10.0 | 1927 | $-1.41^{+0.33}_{-0.27}$ | $-19.23^{+0.71}_{-0.70}$ | $0.69^{+0.25}_{-0.23}$ | $0.89^{+0.42}_{-0.50}$ | $-8.84^{+0.42}_{-0.52}$ |
| | 10.0–10.5 | 1053 | $-1.07^{+0.50}_{-0.37}$ | $-19.48^{+1.15}_{-0.85}$ | $0.91^{+0.38}_{-0.33}$ | $1.21^{+0.45}_{-0.60}$ | $-9.00^{+0.39}_{-0.62}$ |
| | 10.5–11.0 | 501 | $-0.42^{+0.77}_{-0.53}$ | $-18.89^{+1.44}_{-1.27}$ | $1.12^{+0.48}_{-0.41}$ | $1.38^{+0.5}_{-2.30}$ | $-9.31^{+0.48}_{-2.33}$ |
| | 11.0–11.5 | 136 | $0.26^{+0.58}_{-0.58}$ | $-18.61^{+1.16}_{-1.23}$ | $1.28^{+0.63}_{-0.64}$ | $1.34^{+0.80}_{-2.38}$ | $-9.79^{+0.75}_{-2.45}$ |
| | 8.0–8.5 | 191 | $-2.20^{+0.17}_{-0.09}$ | $-18.54^{+0.71}_{-0.51}$ | $0.30^{+0.09}_{-0.09}$ | $0.24^{+0.15}_{-0.18}$ | $-8.13^{+0.13}_{-0.14}$ |
| | 8.5–9.0 | 1295 | $-1.97^{+0.22}_{-0.19}$ | $-18.95^{+0.68}_{-0.65}$ | $0.41^{+0.11}_{-0.11}$ | $0.54^{+0.19}_{-0.20}$ | $-8.25^{+0.15}_{-0.21}$ |
| | 9.0–9.5 | 1716 | $-1.71^{+0.25}_{-0.22}$ | $-19.34^{+0.80}_{-0.80}$ | $0.54^{+0.13}_{-0.14}$ | $0.87^{+0.26}_{-0.31}$ | $-8.36^{+0.24}_{-0.29}$ |
| $z = 3$ | 9.5–10.0 | 1011 | $-1.45^{+0.29}_{-0.22}$ | $-19.79^{+0.74}_{-0.80}$ | $0.69^{+0.16}_{-0.16}$ | $1.20^{+0.36}_{-0.41}$ | $-8.50^{+0.33}_{-0.41}$ |
| | 10.0–10.5 | 453 | $-1.17^{+0.41}_{-0.30}$ | $-20.10^{+0.95}_{-0.77}$ | $0.77^{+0.35}_{-0.29}$ | $1.28^{+0.53}_{-0.93}$ | $-8.92^{+0.54}_{-0.96}$ |
| | 10.5–11.0 | 186 | $-0.74^{+0.57}_{-0.49}$ | $-20.13^{+1.23}_{-1.03}$ | $0.99^{+0.41}_{-0.57}$ | $1.23^{+0.84}_{-1.51}$ | $-9.44^{+0.82}_{-1.55}$ |
| | 11.0–11.5 | 46 | $-0.15^{+0.56}_{-0.42}$ | $-19.84^{+1.29}_{-0.85}$ | $1.32^{+0.78}_{-0.49}$ | $1.72^{+0.83}_{-2.50}$ | $-9.45^{+0.77}_{-2.45}$ |
| | 8.5–9.0 | 166 | $-2.12^{+0.21}_{-0.14}$ | $-19.39^{+0.59}_{-0.58}$ | $0.33^{+0.07}_{-0.06}$ | $0.66^{+0.18}_{-0.23}$ | $-8.21^{+0.14}_{-0.15}$ |
| | 9.0–9.5 | 608 | $-1.83^{+0.22}_{-0.22}$ | $-19.76^{+0.62}_{-0.69}$ | $0.45^{+0.08}_{-0.09}$ | $0.92^{+0.24}_{-0.24}$ | $-8.33^{+0.20}_{-0.24}$ |
| $z = 4$ | 9.5–10.0 | 434 | $-1.55^{+0.28}_{-0.22}$ | $-20.12^{+0.74}_{-0.66}$ | $0.56^{+0.13}_{-0.11}$ | $1.22^{+0.27}_{-0.32}$ | $-8.48^{+0.28}_{-0.35}$ |
| | 10.0–10.5 | 200 | $-1.33^{+0.42}_{-0.27}$ | $-20.43^{+0.77}_{-0.75}$ | $0.61^{+0.27}_{-0.21}$ | $1.10^{+0.61}_{-0.67}$ | $-9.13^{+0.65}_{-0.68}$ |
| | 10.5–11.0 | 55 | $-0.93^{+0.55}_{-0.39}$ | $-20.69^{+1.05}_{-0.70}$ | $0.46^{+0.40}_{-0.46}$ | $0.82^{+1.37}_{-0.90}$ | $-9.83^{+1.34}_{-0.79}$ |

2. *Stellar mass and UV colors.* At fixed redshifts, galaxies with higher stellar masses display redder UV colors than their lower-mass counterparts, suggesting variations in dust content or age. Notably, there is a distinct turnover in the UV slope as a function of UV magnitude at $z = 2$ and $z = 3$ for higher-mass galaxies, accompanied by a wider spread in the $1\sigma$ distribution. This pattern could indicate greater dust attenuation at higher masses, influencing the observed UV brightness and colors.
3. *Dust attenuation.* Across the same stellar masses, higher-redshift galaxies generally show lower levels of dust attenuation, consistent with the observed bluer values of $\beta$. This trend implies an inverse relationship between redshift and dust content within these stellar mass ranges.
4. *Star formation rates.* Star formation rates increase with stellar mass, aligning with the star-forming main sequence (K. G. Noeske et al. 2007). This correlation suggests that higher-mass galaxies experience enhanced gas accretion rates, fueling more vigorous star formation. Interestingly, at the highest stellar masses, we observe a turnover, where more massive galaxies become redder yet have similar SFRs; this is likely due to higher dust levels obscuring the true UV luminosity, with the total SFR being underestimated in this high-extinction regime.

To illustrate these trends, we highlight two example galaxies in Figure 6, which show DENSE BASIS SEDs fit to the

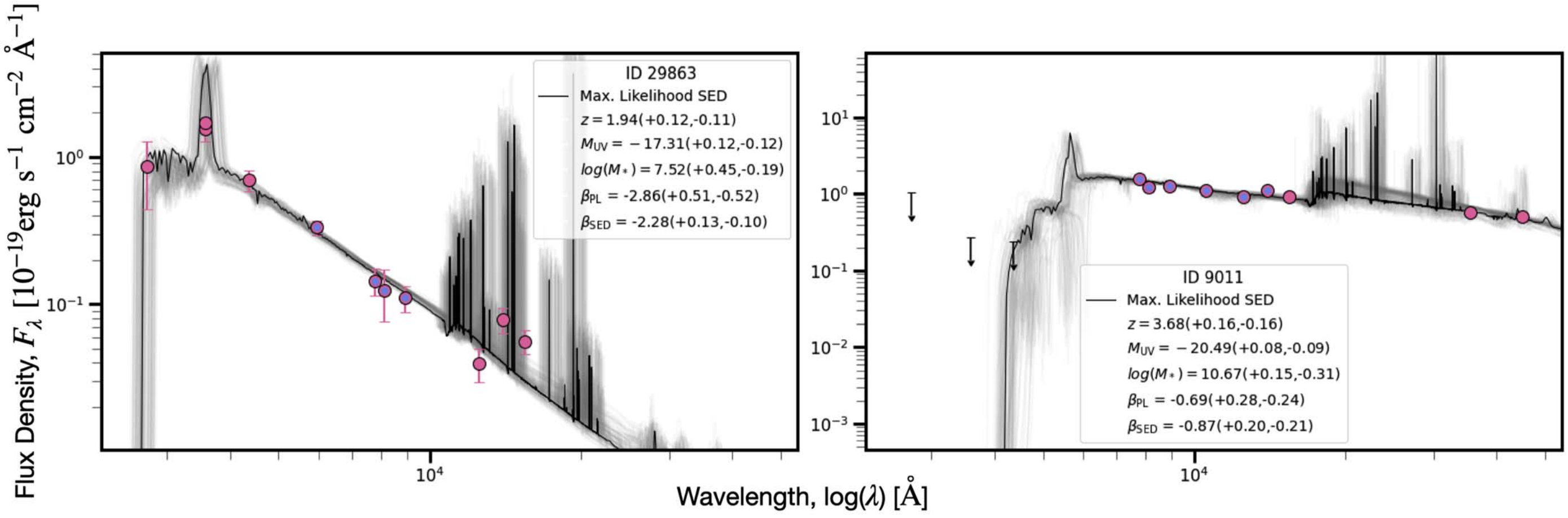

**Figure 6.** Examples of DENSE BASIS SEDs fit to photometric data (pink circles). The photometric data points with blue centers are the data points used to fit $\beta_{\rm PL}$, and the photometric data points with upper-limit arrows indicate low signal-to-noise (SNR $\leqslant$ 2). The full suite of SEDs fit to the data are overplotted in gray, and the maximum-likelihood SED is overplotted in black. These are used to measure the UV slope, $\beta_{\rm SED}$, as well as to obtain median and 1$\sigma$ uncertainties on galaxy properties. Here, we show two cases: The left is a fainter, low-mass source with $M_{\rm UV} = -17.3^{+0.1}_{-0.1}$ mag and $\log[M_*] = 7.5^{+0.5}_{-0.2}\,M_\odot$, and thus a bluer UV spectral slope, $\beta_{\rm SED} = -2.3^{+0.1}_{-0.1}$. The right is a bright, red source with $M_{\rm UV} = -20.5^{+0.1}_{-0.1}$ mag and $\log[M_*] = 10.7^{+0.2}_{-0.3}\,M_\odot$, and thus a redder UV spectral slope, $\beta_{\rm SED} = -0.7^{+0.2}_{-0.2}$.

**Table 5**
Dense Basis Galaxy Parameters and Spearman R Correlation Coefficients for This Sample

| | Parameter | Median and 1$\sigma$ | $\rho$ ($\beta$ vs. Parameter) | $p$-value |
|---|---|---|---|---|
| $z = 2$ | $\beta$ | $-1.63^{+0.51}_{-0.33}$ | … | … |
| | $z$ | $1.98^{+0.31}_{-0.29}$ | $-0.03^{+0.01}_{-0.01}$ | $0.00^{+0.01}_{-0.00}$ |
| | $M_{\rm UV}$ | $-18.54^{+0.89}_{-1.02}$ | $-0.06^{+0.01}_{-0.01}$ | $0.00^{+0.00}_{-0.00}$ |
| | $\log(M_*/M_\odot)$ | $9.14^{+0.84}_{-0.62}$ | $0.57^{+0.01}_{-0.01}$ | $0.00^{+0.00}_{-0.00}$ |
| | $A_{\rm v}$ | $0.59^{+0.27}_{-0.20}$ | $0.42^{+0.01}_{-0.01}$ | $0.00^{+0.00}_{-0.00}$ |
| | log(SFR) | $0.50^{+0.63}_{-0.44}$ | $0.28^{+0.01}_{-0.01}$ | $0.00^{+0.00}_{-0.00}$ |
| | log(sSFR) | $-8.52^{+0.31}_{-0.57}$ | $-0.38^{+0.01}_{-0.01}$ | $0.00^{+0.00}_{-0.00}$ |
| $z = 3$ | $\beta$ | $-1.68^{+0.42}_{-0.35}$ | … | … |
| | $z$ | $2.98^{+0.29}_{-0.31}$ | $0.02^{+0.01}_{-0.01}$ | $0.20^{+0.38}_{-0.16}$ |
| | $M_{\rm UV}$ | $-19.36^{+0.85}_{-0.92}$ | $-0.08^{+0.01}_{-0.01}$ | $0.00^{+0.00}_{-0.00}$ |
| | $\log(M_*/M_\odot)$ | $9.26^{+0.68}_{-0.45}$ | $0.57^{+0.01}_{-0.01}$ | $0.00^{+0.00}_{-0.00}$ |
| | $A_{\rm v}$ | $0.54^{+0.23}_{-0.18}$ | $0.42^{+0.01}_{-0.01}$ | $0.00^{+0.00}_{-0.00}$ |
| | log(SFR) | $0.80^{+0.53}_{-0.41}$ | $0.25^{+0.01}_{-0.01}$ | $0.00^{+0.00}_{-0.00}$ |
| | log(sSFR) | $-8.37^{+0.25}_{-0.43}$ | $-0.38^{+0.01}_{-0.01}$ | $0.00^{+0.00}_{-0.00}$ |
| $z = 4$ | $\beta$ | $-1.69^{+0.39}_{-0.34}$ | … | … |
| | $z$ | $3.85^{+0.39}_{-0.25}$ | $-0.14^{+0.02}_{-0.02}$ | $0.00^{+0.00}_{-0.00}$ |
| | $M_{\rm UV}$ | $-19.92^{+0.73}_{-0.78}$ | $-0.01^{+0.01}_{-0.01}$ | $0.74^{+0.18}_{-0.25}$ |
| | $\log(M_*/M_\odot)$ | $9.47^{+0.61}_{-0.39}$ | $0.52^{+0.02}_{-0.02}$ | $0.00^{+0.00}_{-0.00}$ |
| | $A_{\rm v}$ | $0.48^{+0.17}_{-0.13}$ | $0.33^{+0.02}_{-0.02}$ | $0.00^{+0.00}_{-0.00}$ |
| | log(SFR) | $0.97^{+0.40}_{-0.34}$ | $0.13^{+0.02}_{-0.02}$ | $0.00^{+0.00}_{-0.00}$ |
| | log(sSFR) | $-8.41^{+0.28}_{-0.49}$ | $-0.38^{+0.00}_{-0.01}$ | $0.00^{+0.00}_{-0.00}$ |
| Median of Entire Sample | $\beta$ | $-1.65^{+0.47}_{-0.33}$ | … | … |
| | $z$ | $2.25^{+0.92}_{-0.48}$ | $-0.08^{+0.00}_{-0.00}$ | $0.00^{+0.00}_{-0.00}$ |
| | $M_{\rm UV}$ | $-18.91^{+1.08}_{-1.04}$ | $-0.02^{+0.00}_{-0.00}$ | $0.00^{+0.01}_{-0.00}$ |
| | $\log(M_*/M_\odot)$ | $9.23^{+0.75}_{-0.56}$ | $0.56^{+0.00}_{-0.00}$ | $0.00^{+0.00}_{-0.00}$ |
| | $A_{\rm v}$ | $0.56^{+0.26}_{-0.19}$ | $0.41^{+0.01}_{-0.01}$ | $0.00^{+0.00}_{-0.00}$ |
| | log(SFR) | $0.65^{+0.59}_{-0.49}$ | $0.24^{+0.01}_{-0.01}$ | $0.00^{+0.00}_{-0.00}$ |
| | log(sSFR) | $-8.46^{+0.30}_{-0.55}$ | $-0.38^{+0.00}_{-0.01}$ | $0.00^{+0.00}_{-0.00}$ |

**Note.** Here, the median and uncertainties in $\rho$ and $p$-value take into account individual uncertainties in $\beta$ and each galaxy parameter.

photometry. The bluer source shown in this figure is at redshift $z = 1.94^{+0.12}_{-0.11}$, has a UV magnitude of $M_{\rm UV} = -17.31^{+0.12}_{-0.12}$ mag, and has a stellar mass of $\log[M_*] = 7.52^{+0.51}_{-0.19}\,M_\odot$, corresponding to a significantly bluer UV spectral slope ($\beta_{\rm SED} = -2.28^{+0.13}_{-0.10}/\beta_{\rm PL} = -2.86^{+0.51}_{-0.52}$). In contrast, the edder source, at redshift $z = 3.68^{+0.16}_{-0.16}$, is brighter ($M_{\rm UV} = -20.49^{+0.08}_{-0.09}$ mag) and more massive ($\log[M_*] = 10.67^{+0.15}_{-0.31}\,M_\odot$), resulting in a redder UV spectral slope ($\beta_{\rm SED} = -0.87^{+0.28}_{-0.24}/\beta_{\rm PL} = -0.69^{+0.28}_{-0.24}$). These galaxies depict the trends observed in our sample: lower-mass galaxies tend to be bluer, fainter, and have lower dust attenuation, while higher-mass galaxies are redder, brighter, and dustier. This comparison reinforces the physical connections between UV spectral slopes, stellar mass, and other galaxy properties, as discussed earlier.

## 4. Discussion

### 4.1. Modeling Caveats

In employing the DENSE BASIS framework for constructing our sample's SEDs, we adopt a set of priors that accommodate a broad spectrum of star-forming galaxy SED models. These priors, which are summarized in Table 3, include a wide range of metallicity from −2.0 to 0.25 in units of $\log(Z_\odot)$, a dust attenuation ($A_{\rm v}$) spanning from 0.0 to 4.0 mag, and a log-scale specific star formation rate (sSFR) from −14.0 to −7.0 per year. These settings allow for extensive variability in galaxy properties, but they also introduce certain constraints that could impact the reliability of our model SEDs.

While the default stellar grids used in DENSE BASIS provide a solid foundation, they are not without limitations. For instance, the stellar grids can lead to a $\beta$ "floor," where the UV slope $\beta$ does not realistically extend below approximately −2.5, potentially skewing the SED fits for galaxies with extreme properties. This is a critical consideration, as it may bias our interpretations of the youngest and lowest-metallicity galaxies.

To potentially enhance the accuracy and realism of our SED models, future work could explore the incorporation of alternative stellar models, such as those provided by E. Zackrisson et al. (2011). These models offer variability in the IMF, metallicity, and star formation history, which could address the biases introduced by the current limitations of our stellar grids. Incorporating these bluer stellar models could significantly alleviate the $\beta$ "floor" effect by extending the range of modeled UV slopes.

Despite these potential enhancements, it is important to note that the current sample's median and $1\sigma$ spread in $\beta$ across all redshifts sits well above the artificial floor, indicating that, for the majority of our data set, the existing models suffice without significant bias. Consequently, while acknowledging these limitations, we recommend the exploration of expanded stellar grid models in future studies.

### *4.2. Comparisons with Previous Work*

In this work, we directly compare our results to the most recent works within the same redshift range. We note, however, that this analysis will only compare general trends in $\beta$, $M_{\rm UV}$, and stellar mass because UV slope measurements and measurements of galaxy properties are sensitive to sample selection and methodology, including but not limited to redshift, magnitude, stellar mass, and the wavelength ranges used to measure the UV slope.

We first compare the results for $\beta$, $M_{\rm UV}$, and stellar mass for our sample with that of N. P. Hathi et al. (2013), due to the overlapping nature of our investigations in the GOODS-S field and the corresponding redshift range. They note that their range of UV magnitudes at average redshifts $<z> = 1.6, 2.2, 2.6, 3.8$ (when a source number cut of $N > 40$ is implemented) spans $-23 < M_{\rm UV} < -19$, which is slightly brighter than what our sample ranges with $N > 20$ spanning $-21 < M_{\rm UV} < -17$. Despite these disparities, both of our median stellar masses remain consistent, averaging around $\sim 10^9\, M_\odot$ for the entire sample. Any differences in $M_{\rm UV}$ could be due to selection criteria and/or the subsequent SED fitting to the observational data, where the values we utilize are measured from rest wavelengths defined in D. Calzetti et al. (1994; for this analysis, we quote their values for their fits with LEPHARE; see their Table 1). In terms of $\beta$, our study yields a slightly redder average $\beta_{\rm SED} \sim -1.65^{+0.47}_{-0.33}$, encompassing a wider range of values, compared to Hathi et al.'s average of $\beta_{\rm SED} \sim -1.72 \pm 0.38$. This discrepancy could potentially be attributed to the specific region of the SED utilized to measure $\beta$ (they measure $\beta$ from $\lambda_{\rm rest} = 1500$–$1900$ Å), as well as differences in sample size and the focus on a single CANDELS field in their study.

We also compare our average values of $\beta$ to those of N. P. Hathi et al. (2016) at $z = 2$–$2.5$ from the VIMOS Ultra-Deep Survey (VUDS), which utilizes deep photometric data in three fields (ECDFS, VVDS, and COSMOS). Their sample takes the average $\beta_{\rm phot}$ from 854 faint star-forming galaxies whose UV slopes are measured directly from the photometric bands around $\lambda_{\rm rest} = 1200$–$3000$ Å and average $\beta_{\rm phot} = -1.32 \pm 0.02$. Their sample spans $-22 < M_{\rm UV} < -19$ and stellar masses span $9.0 < \log(M_\odot) < 11.0$ (average $\sim 10^{9.5}\, M_\odot$) at $z \sim 2$. The observed marginal increase in brightness and average stellar mass of these sources can be attributed to the survey limitations of VUDS, which targets star-forming galaxies brighter than $i_{\rm AB} \sim 25$ mag, potentially influencing the completeness of trends in $M_{1500} - \beta$ and $M_* - \beta$ relations; however, the data still reveal general trends, with $M_{1500} - \beta$ showing little to no trend, while $M_* - \beta$ demonstrates the expected trend of a redder $\beta$ with increasing mass.

At $z = 4$, we compare our work to that of R. J. Bouwens et al. (2014), whose sample came from galaxies in the HST HUDF, ERS, CANDELS-N, and CANDELS-S surveys. With a larger sample, their work was able to span $-22 < M_{\rm UV} < -16$ with a weighted mean $\beta \sim -1.98$, which is slightly bluer than our average of $\beta_{\rm SED} = -1.69$ at $z = 4$. It is possible that variations in the average $\beta$ come from how they define it in their work—where they fit a power law to the $J_{125}$-band and $H_{160}$-band fluxes (e.g., see their Equation 1). Their analysis also makes use of deeper data (e.g., the HUDF), leading to a fainter average magnitude. They also observe a correlation between $M_{\rm UV}$ and $\beta$. However, they define UV magnitude at $\lambda_{\rm rest} = 2000$ Å, and they state that works that define it at $\lambda_{\rm rest} = 1500$ Å (such as ours) may not see any correlations. This bears out, as we see no significant correlation (a Spearman coefficient of $\rho = -0.02$($1\sigma$CI: [–0.04, $-0.01$])). As described in Section 3, we observe a much stronger correlation between $\beta$ and stellar mass. A comparison of these two approaches to measuring the UV magnitude is beyond the scope of this work, but is left to future works.

The work of P. Kurczynski et al. (2014) examines galaxies at $z \sim 1$–$8$ in the UVUDF field, whose $M_{2330}$ goes down to $\sim -14$. When comparing our average values to theirs, we sample their biweighted means for $\beta$ and corresponding errors (given in Table 1 of P. Kurczynski et al. 2014). Given that their work has a majority of sources within $-16 < M_{2330} < -18$, mostly fainter than our sample, as a function of redshift, their sample is on average bluer than ours, with the expectation that these are, on average, lower-mass galaxies. Given the fact that they measure UV magnitude at 2330 Å (as discussed above), they state that they find some trends in $\beta$ and UV magnitude, where $d\beta/dM = -0.11 \pm 0.01$ (random) $\pm 0.1$ (systematic), which points to less luminous galaxies being bluer in their UV colors.

The faintest objects in the P. Kurczynski et al. (2014) sample are at $z = 1$ and are likely due to the increased sensitivity and availability of data at lower redshifts. Their $z = 2$–$4$ sample is comparable to ours, with one object slightly fainter than $M_{\rm UV} = -16$, while our work extends down to $M_{\rm UV} = -16$. R. J. Bouwens et al. (2014) reports comparable magnitudes in their sample to those in our work. However, both R. J. Bouwens et al. (2014) and P. Kurczynski et al. (2014) also make use of the deeper HUDF data, which provide greater sensitivity compared to the CANDELS/UVCANDELS photometry used in our work, which does not overlap with the HUDF. As discussed in R. J. Bouwens et al. (2014), the HUDF fields are smaller but significantly deeper in every filter band, enabling the detection of fainter galaxies. This is further emphasized in Section 4 of P. Kurczynski et al. (2014), where they state that the HUDF data detect galaxies down to an absolute magnitude of $M_{\rm UV} = -14$.

Finally, we compare our results at $z = 4$ to those of S. L. Finkelstein et al. (2012), who gave average values for the UV slope for their sample from $z = 4$–$8$ in the GOODS-S field, and also discussed significant correlations between $\beta$ and stellar mass, which are similar to what we observe here. Their average $\beta$ for their entire sample at $z = 4$ is $\beta = -1.82$, which is slightly bluer than our average $\beta_{\rm SED}$ at this redshift. However, their median stellar mass values lie at $\log(M_*) \sim 9.0\, M_\odot$, while this work is averaged at $\log(M_*) \sim 9.5\, M_\odot$. Their data also made use of a smaller area, including the deeper HUDF, so they would have a reduced sensitivity to more massive, redder galaxies compared to our sample. When binned by stellar mass, our average UV slope aligns most with galaxies at $\log(M_*) = 9$–$10\, M_\odot$—the same is true for their analysis. Despite employing different SED-fitting software to derive the UV slope and related galaxy properties, the consistent finding of a correlation between $\beta$ and stellar mass across both studies suggests that this relationship is robust, transcending the specifics of the computational tools used.

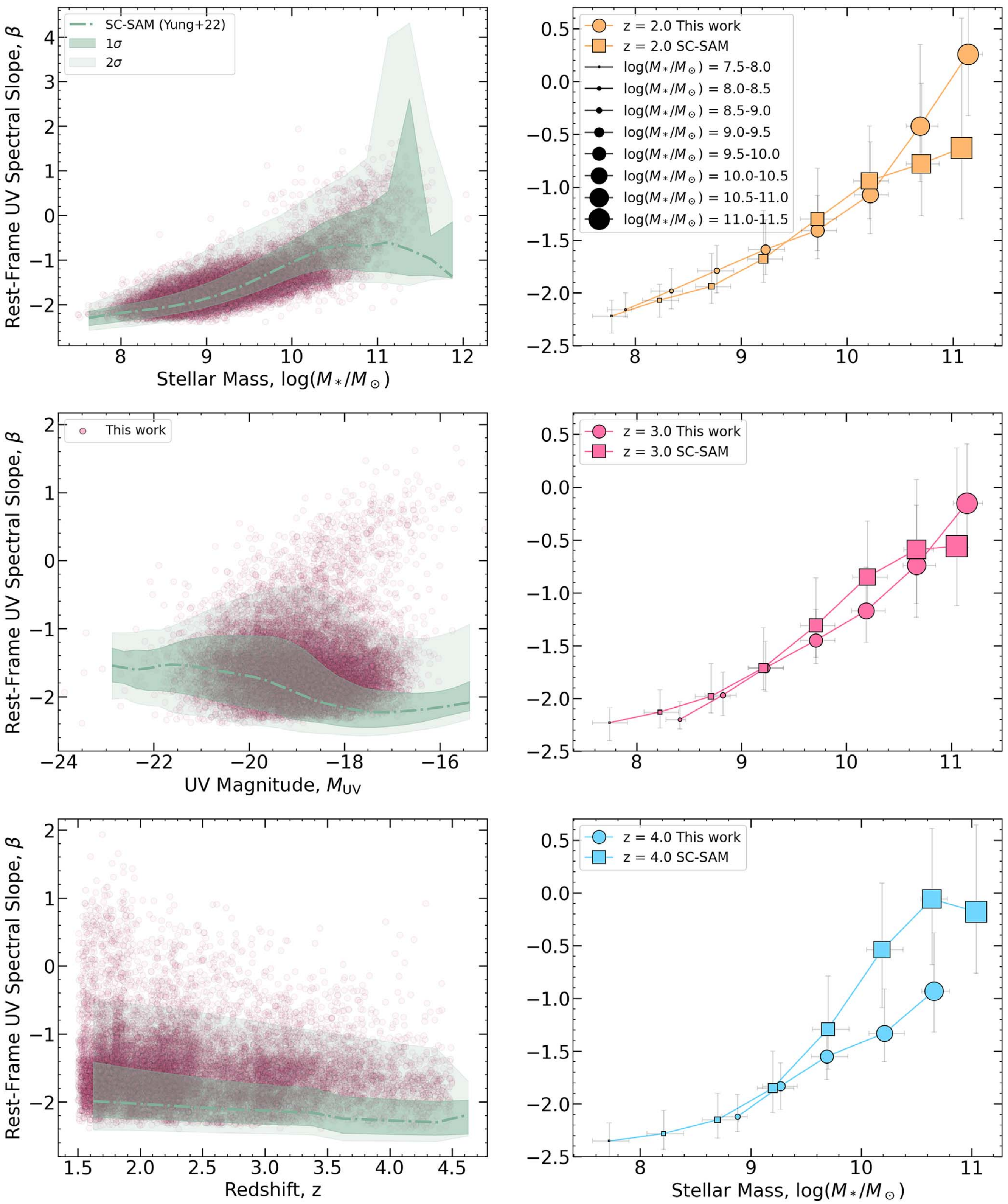

**Figure 7.** Left: comparison of our work for $\beta_{\mathrm{SED}}$ (pink scatter points) to SC-SAM theoretical predictions (R. S. Somerville et al. 2021; L. Y. A. Yung et al. 2022). The green track marks the median, and the shaded area marks the $1\sigma$ and $2\sigma$ regions for galaxies in bins of stellar mass (top), UV magnitude (middle), and redshift (bottom). On average, we see that simulations mostly agree with $\beta$ values, where brighter, more massive sources exhibit redder UV colors. As a function of redshift, simulations also trend toward slightly bluer UV colors. Right: comparisons of the median UV spectral slope and $1\sigma$ error bars for observed (circles) and simulated (squares) galaxies in bins of $N > 20$ as a function of stellar mass and redshift. We show that simulations and observations for each mass bin and redshift are in good agreement. However, at high stellar masses, namely at $z = 4$, the SC-SAM galaxies are redder than this sample, in which strong dust attenuation is required in order for the simulated galaxy populations to match observed UV LF constraints (S. L. Finkelstein 2016).

### 4.3. Comparing Observations to Simulations

Here, we compare the results of our observations with predictions from the Santa Cruz semi-analytic model (SC-SAM) JWST wide-field catalog (R. S. Somerville et al. 2021; L. Y. A. Yung et al. 2022). From these catalogs, we pull sources from each of the four CANDELS fields used in this work from each of the eight realizations the catalogs offer, along with various galaxy properties (stellar mass, $M_{1500}$, $M_{2300}$, and redshift). The galaxies in the mock lightcones are simulated with the SC-SAM for galaxy formation (R. S. Somerville et al. 2015). The free parameters in the model are calibrated to reproduce a set of galaxy properties observed at $z \sim 0$ and have been shown to well reproduce the observed evolution of one-point distribution functions of $M_{\rm UV}$, $M_*$, and SFR (R. S. Somerville et al. 2015; L. Y. A. Yung et al. 2019a, 2019b) over a wide range of redshift. In the modeling of stellar populations and dust attenuation, as summarized in Section 2.5 of L. Y. A. Yung et al. (2019a), galaxies are represented on a 2D grid, with mass distributions of stars organized by age and metallicity. Stellar population synthesis models by G. Bruzual & S. Charlot (2003) are employed to generate unattenuated synthetic SEDs for each galaxy. Dust attenuation is modeled using a simple "slab" approach with D. Calzetti et al. (2000) attenuation curves and assumes a face-on extinction optical depth in the $V$ band, which accounts for the galaxy's inclination, the metallicity of the cold gas, and the cold gas mass and radius. This approach is fine-tuned with parameters adjusted to better match observations, in particular in terms of high-redshift galaxy characteristics.

$\beta$ is calculated from the difference in $M_{1500}$ and $M_{2300}$ by the following relation: $\beta_{\rm UV} = -0.4(M_{\rm FUV} - M_{\rm NUV})/\log(\lambda_{\rm FUV} - \lambda_{\rm NUV}) - 2$ with magnitude in far-UV (FUV, centered at 1500 Å) and near-UV (NUV, centered at 2300 Å) bands. This comparison is similar to what was done in A. M. Morales et al. (2024), where their SC-SAM galaxies with GUREFT merger trees have UV slopes that are defined with the same equation (L. Y. A. Yung et al. 2024). Given that we did not have full access to the synthetic photometry for the SC-SAM galaxies to fit with DENSE BASIS (which would be computationally expensive regardless), measuring UV slopes directly from UV magnitudes provides a reasonable first-order estimate for these models. Here, we only keep sources that have redshifts between $1.5 \leqslant z \leqslant 4.5$, stellar masses between $7.5 \leqslant \log(M_*) \leqslant 12.0$, and dust-corrected UV magnitudes $-23 < M_{1500} < -15$. These selection criteria leaves us with about 16 million simulated central galaxies to compare to observations.

In Figure 7, we plot our full sample versus the median, $1\sigma$, and $2\sigma$ regions of the simulation's distribution. Our analysis indicates that, while general trends in $\beta$ with respect to stellar mass and redshift are consistent between SC-SAM simulations and our observations, notable deviations exist, in particular in specific bins. At redshift $z = 2$, our observed median $\beta$ is $-1.65^{+0.47}_{-0.33}$, which is less steep compared to SC-SAM's $\beta = -2.13$, suggesting that our sample may capture a less extreme range of UV colors, potentially due to differences in galaxy populations or observational constraints. Similarly, at $z = 4$, the SC-SAM predicts redder slopes for higher-mass galaxies ($\log[M_*] > 10.5$) than we observe, suggesting that the models may overestimate dust effects. This discrepancy is apparent as the simulated galaxies at high stellar masses exhibit UV colors that are significantly redder than those observed, necessitating substantial dust attenuation in the simulations to align with observed UV luminosity function (LF) constraints (see S. L. Finkelstein 2016, and references therein). The trends toward bluer UV slopes at higher redshifts and lower stellar masses are consistent among both data sets. We note that all four fields times eight realizations of the SC-SAM cover a total of 32,816 arcmin$^2$. This might include galaxies that are too rare to be found in the UVCANDELS sample. We refer the reader to L. Y. A. Yung et al. (2019a) for further discussion. Further investigations are warranted to explore these discrepancies, and enhancements in model sophistication or expanding observational data sets to include a wider range of galaxy environments may help reconcile these differences.

## 5. Conclusions

Using data from the HST UVCANDELS survey, we analyze the evolution of the rest-frame UV spectral slope as a function of redshift from $z \sim 2$–4. We measure the UV spectral slope both via SED fitting with DENSE BASIS and via photometric power-law fitting to the observed photometry. We compare our observed UV slopes to the measured stellar mass, UV magnitude, redshift, SFR, dust attenuation, and sSFR. With these comparisons, we reach the following conclusions:

1. We find a reasonable agreement between the SED-fitting and power-law-fitting methodologies for the UV spectral slope, though the latter exhibits slightly more significant scatter and preferentially bluer colors on average. This is a limitation of the power-law fitting wherein far fewer data points are utilized to obtain $\beta$ when compared to SED fitting, given that the SED is a good fit to the photometry, though the SED-fitting method is also limited based on the stellar templates used.
2. When measuring $\beta$ via the SED-fitting method for the entire sample, we obtain an average UV slope$\beta_{\rm SED} = -1.65^{+0.47}_{-0.33}$, compared to the $\beta_{\rm PL} = -1.72^{+0.58}_{-0.46}$ obtained via photometric power-law fitting. These average values indicate that the galaxies in our sample exhibit signs of moderate dust attenuation and stellar masses and are quite active, given their specific star formation rates. These results are consistent with previous studies (e.g., N. P. Hathi et al. 2013) covering similar redshift ranges.
3. We compare our results for the sample as a whole and when binned as a function of stellar mass and redshift. We show that, across all three redshift ranges, galaxies with lower stellar masses exhibit bluer UV colors than their counterparts with higher stellar masses, and these trends extend into trends in $\beta$ versus other galaxy properties. Lower-mass galaxies have less dust attenuation, lower star formation rates, and higher specific star formation rates than higher-mass galaxies. As a function of redshift, we find that, on average, galaxies of the same stellar mass but increasing redshift will exhibit bluer UV colors and brighter UV magnitudes.
4. We compare our observations with the results of the SC-SAM simulations (R. S. Somerville et al. 2021; L. Y. A. Yung et al. 2022) and find that the trends in their sources generally agree with our observations in terms of average UV slope and corresponding galaxy property from observations at these redshifts and in these fields, yet their median SC-SAM values are slightly bluer than those of our sample. Although, as we show at the high-stellar-mass end, simulated galaxies are exhibiting redder UV colors at $z = 4$ as a result of dust.

Moving forward, refining the integration of SED-fitting techniques and observational data will be crucial in further dissecting the intrinsic properties of galaxies, in particular in the context of evolving redshift and stellar mass. Future efforts could focus on expanding the range of stellar templates and exploring alternative dust models to better capture the observed UV colors of galaxies, allowing us to infer the complexities of galaxy evolution. Additionally, increased observational coverage and deeper field surveys could provide a richer data set for validating and enhancing simulations. This continuous improvement in both observational strategies and theoretical modeling will be vital for a more comprehensive understanding of galaxy properties across different epochs of the Universe.

## Acknowledgments

A.M.M. and S.L.F. acknowledge support from NASA through STScI HST GO award GO-15647. A.M.M. acknowledges support from the National Science Foundation Graduate Research Fellowship Program under grant No. DGE 2137420. Any opinions, findings, conclusions, or recommendations expressed in this material are those of the author(s) and do not necessarily reflect the views of the National Science Foundation. R.A.W. acknowledges support from NASA JWST Interdisciplinary Scientist grant Nos. NAG5-12460, NNX14AN10G, and 80NSSC18K0200 from GSFC.

*Software*: IPython (F. Pérez & B. E. Granger 2007), matplotlib (J. D. Hunter 2007), NumPy (S. Van Der Walt et al. 2011), SciPy (T. E. Oliphant 2007), Astropy (T. P. Robitaille et al. 2013), Dense Basis (K. G. Iyer et al. 2019), emcee (D. Foreman-Mackey et al. 2013).

## ORCID iDs

Alexa M. Morales https://orcid.org/0000-0003-4965-0402
Steven L. Finkelstein https://orcid.org/0000-0001-8519-1130
Micaela B. Bagley https://orcid.org/0000-0002-9921-9218
Anahita Alavi https://orcid.org/0000-0002-8630-6435
Norman A. Grogin https://orcid.org/0000-0001-9440-8872
Nimish P. Hathi https://orcid.org/0000-0001-6145-5090
Anton M. Koekemoer https://orcid.org/0000-0002-6610-2048
Kalina V. Nedkova https://orcid.org/0000-0001-5294-8002
Laura Prichard https://orcid.org/0000-0002-0604-654X
Marc Rafelski https://orcid.org/0000-0002-9946-4731
Ben Sunnquist https://orcid.org/0000-0003-3759-8707
Sina Taamoli https://orcid.org/0000-0003-0749-4667
Harry I. Teplitz https://orcid.org/0000-0002-7064-5424
Xin Wang https://orcid.org/0000-0002-9373-3865
Rogier A. Windhorst https://orcid.org/0000-0001-8156-6281
L. Y. Aaron Yung https://orcid.org/0000-0003-3466-035X